\documentclass[preprint2]{aastex}
\usepackage{graphicx}
\usepackage{physics}

\begin{document}

\title{Excitation of Langmuir waves in the magnetospheres of AGN}


\author{Z.N. Osmanov}

\affil{School of Physics, Free University of Tbilisi, 0183, Tbilisi,
Georgia, Email: z.osmanov@freeuni.edu.ge}
\affil{E. Kharadze Georgian National Astrophysical Observatory, Abastumani, 0301, Georgia}

\begin{abstract}
In the paper we study the process of excitation of Langmuir waves in the magnetospheres of active galactic nuclei (AGN), by taking a general-relativistic expression of the Goldreich-Julian density into account. We considered the linearised set of equations which describe dynamics of the studied mechanism: the Euler equation, the continuity equation and the Poisson equation. After solving the dispersion relation and obtaining the instability growth rate, we explored it versus several physical parameters: electron's and proton's relativistic factors and the mass and luminosity of AGN, which are supposed to be Kerr black holes. We showed that the parametric process of energy pumping into the Langmuir waves is very efficient and the electrostatic field's amplitude will be exponentially amplifying, which might account for pair creation, particle acceleration and plasma heating processes in the nearby regions of AGN.

\end{abstract}


\section{Introduction} 

One of the fundamental problems of astrophysics of compact objects is understanding of a particular mechanism responsible for energy pumping into waves, very high energy (VHE) particles and radiation. Very often, all these three might be closely related to each other. In particular, for the magnetospheres of pulsars \citep{review,sc1,sc2} and black holes \citep{heat,pev,stellar,zev} it was shown that particles might be energized to VHE via the magneticentrifugally excited Langmuir waves.

The role of centrifugal effects in the rotating magnetospheres has been considered by \cite{gold1,gold2} for pulsars, where the author has pointed out that for accelerating particles to high energies two necessary factors should exist: strong magnetic field and rotation. 

It is strongly believed that the nearby regions of AGNs are characterized by magnetic fields, strong enough to bound the charged particles on the field lines - frozen in condition \citep{BZ}. On the other hand, magnetospheres of AGN are rotating, because of the central object which is assumed to be a Kerr-type black hole \citep{carroll,shapiro}. Therefore, the magnetocentrifugal effects should be essential.

Originally the idea of the centrifugally driven electrostatic field for the black hole magnetospheres has been proposed by \cite{lang} where it has been found that due to the frozen-in condition the plasma particles follow the co-rotating magnetic field lines, experience centrifugal force, which parametrically drives the Langmuir waves via charge separation. During this process, the electric field's amplitude will evolve with time exponentially and under certain conditions, it might reach a threshold value $1.4\times10^{14}$ statvolt cm$^{-1}$, when efficient pair production will start via the Schwinger process \citep{schwinger}.

In the aforementioned paper a simple expression of particles' density has been used. In particular, the so-called Goldreich-Julian (GJ) density \citep{GJ} without taking general relativistic effects into account has been considered. On the other hand, general-relativistic corrections might be important. This problem has been studied by \cite{rieger} where the authors generalized an expression of the GJ denisity for Kerr-type black holes.

Here extend our previous works for more general case and study how the Langmuir modes are driven in the AGN magnetospheres.

The paper is organized  as follows: in Sec.2, general approach is considered, in Sec. 3, the model is applied to an AGN environment, for which we obtain results, and in Sec. 4, we sumarize them. 

\section{Main consideration}

According to the original concept, the generation of Langmuir waves in a medium rotating with the angular velocity, $\Omega_0$, is described by the system of linearized equations (in Fourier space) \citep{heat,pev,zev} 
 \begin{equation}
\label{eul} \frac{\partial p_{_{\beta}}}{\partial
t}+ik\upsilon_{_{\beta0}}p_{_{\beta}}=
\upsilon_{_{\beta0}}\Omega_0^2r_{_{\beta}}p_{_{\beta}}+\frac{e_{_{\beta}}}{m_{_{\beta}}}E,
\end{equation}
\begin{equation}
\label{cont} \frac{\partial n_{_{\beta}}}{\partial
t}+ik\upsilon_{_{\beta0}}n_{_{\beta}}, +
ikn_{_{\beta0}}\upsilon_{_{\beta}}=0
\end{equation}
\begin{equation}
\label{pois} ikE=4\pi\sum_{_{\beta}}n_{_{\beta0}}e_{_{\beta}},
\end{equation}
where Eq. (\ref{eul}) is the equation of motion by taking a centrifugal force into account, Eq. (\ref{cont}) is the continuity equation, and Eq. (\ref{pois}) is the Poisson equation. Here,
 ${\beta}$ denotes an index of species (protons or electrons), $p_{_{\beta}}$ is the dimensionless first order momentum, $k$ is the wave number of the corresponding wave, $\upsilon_{_{\beta0}}(t) \approx
c\cos\left(\Omega_0 t + \phi_{_{\beta}}\right)$ denotes the unperturbed
velocity and $r_{_{\beta}}(t) \approx
\frac{c}{\Omega_0}\sin\left(\Omega_0 t + \phi_{_{\beta}}\right)$ is the
radial coordinate, $e_{_{\beta}}$ is the particle's
charge, $n_{_{\beta}}$ and $n_{_{\beta0}}$ are the first and zeroth order Fourier terms of the number density. 

As it is clear from the first term of the right hand side of Eq. (\ref{eul}), which is the contribution from the centrifugal force, it differenciates between electrons and protons, which leads to charge separation. This in turn, via the Poisson equation, excites the electrostatic waves. By following the aforementioned papers and substituting the ansatz 
\begin{equation}
\label{ansatz}
n_{\beta}=N_{\beta}e^{-\frac{iV_{\beta}k}{\Omega_0}\sin\left(\Omega_0 t
+ \phi_{\beta}\right)},
\end{equation}
into Eqs. (\ref{eul}-\ref{pois}), one can derive the dispersion relation of the induced modes
\begin{equation}
\label{disp} \omega^2 -\omega_e^2 - \omega_p^2  J_0^2(b)= \omega_p^2
\sum_{\mu} J_{\mu}^{2}(b) \frac{\omega^2}{(\omega-\mu\Omega_0)^2},
\end{equation}
where $\omega$ is the frequency of the Langmuir wave, $\omega_{e,p}\equiv\sqrt{4\pi e^2n_{e,p}/m_{e,p}\gamma_{e,p}^3}$ denotes the Langmuir frequency of the corresponding component, $n_{e,p}$, $m_{e,p}$ and $\gamma_{e,p}$ are the charge, mass and the relativistic factor of electrons and protons respectively, $J_{\mu}(x)$ denotes the Bessel function of the first kind, $b=2ck\sin\phi/\Omega_0$ and $\phi$ is a phase parameter. It was found by \cite{heat,pev,zev} that the aforementioned equation characterised by resonance conditions, $\omega = \mu\Omega_0+\delta$ with $\delta<<\omega_e$, leads to the following cubic equation
\begin{equation}
 \label{disp1}
 \Delta^3=\frac{\omega_e {\omega_p}^2 {J_{\mu}(b)}^2}{2},
 \end{equation}
 having the solution  
\begin{equation}
 \label{grow1}
 \Gamma= \frac{\sqrt3}{2}\left (\frac{\omega_e {\omega_p}^2}{2}\right)^{\frac{1}{3}}
 {J_{\mu}(b)}^{\frac{2}{3}},
\end{equation}
which is the growth rate of the instability, were $\mu = \omega_e/\Omega_0$. One should note that the instability has a parametric character because the relativistic centrifugal force is harmonically time dependent and as a result the amplitude of the electric field will asymptotically increase.

\section{Discusion and results}

A supermassive black hole located in the center of AGN has the angular velocity \citep{carroll}
\begin{equation}
\label{rotat} \Omega_0\approx\frac{a c^3}{GM}\approx
10^{-3}\frac{a}{M_8}rad/s^2,
\end{equation}
which is essential for magnetocentrifugal acceleration. Here, $a$ denotes a dimensionless spin parameter (throughout the paper we consider black holes with $a = 0.1$), $G$ is the gravitational constant, $M$ is the black hole's mass, $M_8\equiv M/(10^8M_{s})$ and $M_{s}\simeq 2\times 10^{33}$g is the solar mass. 

\begin{figure}
  \resizebox{\hsize}{!}{\includegraphics[angle=0]{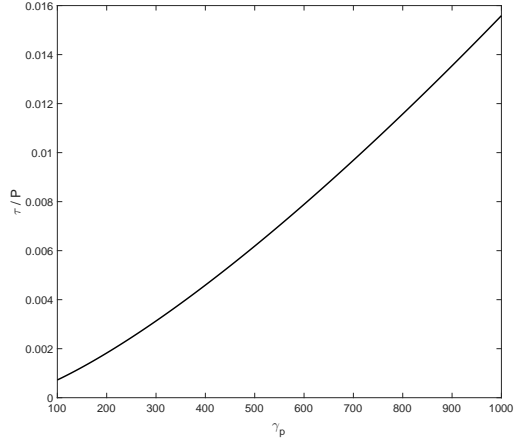}}
  \caption{Behaviour of the dimensionless time-scale versus $\gamma_p$. The parameters are: $M_8 = 1$, $L_{42}=1$, $\theta = \pi/4$, $r = R_{lc}/\sin\theta$, $\gamma_e = 10^2$.}\label{fig1}
\end{figure}

\begin{figure}
  \resizebox{\hsize}{!}{\includegraphics[angle=0]{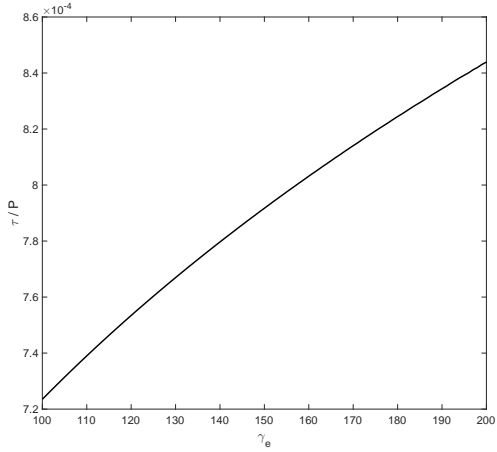}}
  \caption{Behaviour of the dimensionless time-scale versus $\gamma_e$. The parameters are the same as in Fig. 1, except the fixed value of $\gamma_p = 10^2$ and an interval of $\gamma_e = (1-2)\times 10^2$.}\label{fig2}
\end{figure}

\begin{figure}
  \resizebox{\hsize}{!}{\includegraphics[angle=0]{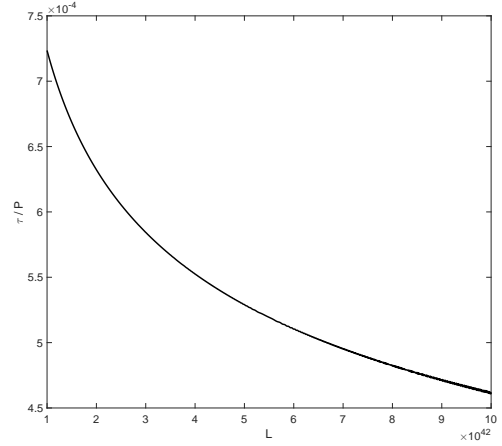}}
  \caption{Dependence of $\tau/P$ on the luminosity of the AGN. The parameters are the same as in Fig. 1, except $\gamma_e = 10^2$ and $L = (1-10)\times 10^{42}$ erg s$^{-1}$.}\label{fig3}
\end{figure}

\begin{figure}
  \resizebox{\hsize}{!}{\includegraphics[angle=0]{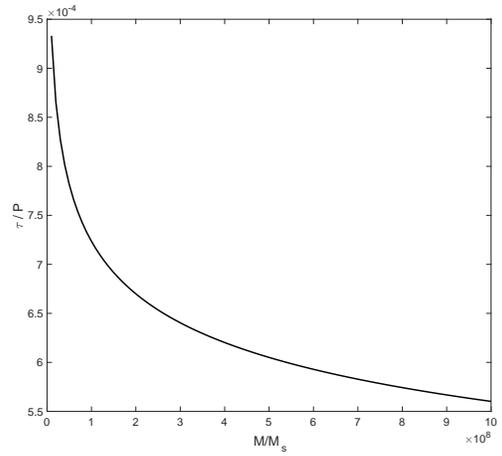}}
  \caption{Dependence of $\tau/P$ on the luminosity of the AGN. The parameters are the same as in the previous figure, except $L = 10^{42}$ erg s$^{-1}$ and $M_8 = 0.1-10$.}\label{fig4}
\end{figure}

Another important factor the present mechanism strongly depends on is the magnetic field. By taking into account the equipartition model (the energy density of magnetic field is of the same order as the energy density of radiation) leads to a following expression
\begin{equation}
 \label{B}
B\simeq \frac{1}{r}\sqrt{\frac{2L}{c}}\simeq 2.8\times 10^2\times \frac{R_H}{r}\times L_{42}^{1/2},
 \end{equation}
 where $R_H\simeq 2GM/c^2$ is the event horizon radius for a small spin parameter, $L$ is the luminosity of AGN and $L_{42}=L/(10^{42} erg\; s^{-1})$.

In the light cylinder zone (a hypothetical area where the linear velocity of rotation equals the speed of light) the particles might be accelerated up to the maximum relativistic factors $\gamma_p^m\simeq\gamma_0\left(eB_{lc}/(2m_p\Omega c)\right)^{2/3}\simeq 1.2\times 10^7\times L_{42}^{1/3}$ \citep{rm}. This is provided by the breakdown of the bead on the wire approximation, when energy of particles is so high that particles do not follow the field lines any more. The acceleration of electrons, in turn, will be limited by means of the inverse Compton mechanism \citep{orb}, $\gamma_e^m\simeq\left(8\pi m_ec^4/(\gamma_0\sigma_TL\Omega)\right)^2\simeq 2.4\times 10^4\times L_{42}^{-2}$. We have assumed that electrons and protons initially are mildly relativistic and have equal relativistic factors of the order of $10$, which is the typical value for AGN outflows.

It is natural to think that rotation can influence the distribution of particles in the magnetosphere, which actually is the GJ density and has originally been introduced for pulsars, but have been used also to AGNs.  \cite{rieger}, considering the gap-type particle acceleration, have generalized the GJ density for the Kerr-type black holes, which for the leading term writes as
\begin{equation}
\label{GJ} 
n_{_{GJ}}\simeq \frac{\left(\Omega-\Omega^F\right)B_Hr_H^2\cos\theta}{\pi ce\alpha_l\rho^2},
\end{equation}
where $\theta$ denotes the angle with respect to the polar axis, $\Omega = 2c\alpha_s r_gr/\Sigma^2$ is the angular velocity relative to absolute space, $\alpha_s = a r_g$, $r_g = GM/c^2$ denotes the gravitational radius, $\Omega^F = c\alpha_s/4r_gr_H$ is the angular velocity of magnetic field lines, $r_H = r_g+\sqrt{r_g^2-\alpha_s^2}$ denotes the radius of the black hole's event horizon, $\alpha_l = \rho\sqrt{\Delta}/\Sigma$, $\rho^2 = r^2+\alpha_s^2\cos^2\theta$, $\Delta = r^2-2rr_g+\alpha_s^2$ and $\Sigma^2 = \left(r^2+\alpha_s^2\right)^2-\alpha_s^2\Delta\sin^2\theta$. For more details please see the work \citep{rieger}.

Now we have to consider the parametric instability of the Langmuir wave excitation and study its efficiency. In the framework of the paper we assume that equipartition energy distribution is valid for each of the species (electrons and protons): $n_{_{GJ}}\gamma_e^m\simeq\gamma_en_e$ and $n_{_{GJ}}\gamma_p^m\simeq\gamma_pn_p$.

As a first example we study the instability growth rate in the light cylinder area, having the radius, $R_l = c/\Omega_0$, versus $\gamma_p$, for the fixed value of $\gamma_e = 10^4$. In Fig. 1 the dependence of the  time-scale of the instability, $\tau = 1/\Gamma$, (normalised by the kinematic time-scale, $P = 2\pi/\Omega_0$, which is the period of rotation of the black hole) is shown. Other parameters are: $M_8 = 1$, $L_{42}=1$, $\theta = \pi/4$, $r = R_{lc}/\sin\theta$, $\gamma_e = 10^2$. As it is evident, the instability time-scale is small compared to the kinematic time-scale, which indicates that the energy pumping mechanism is very efficient, and is characterised by the exponential amplification of the amplitude of electrostatic field. But one should note that $\tau/P$ is an increasing function of $\gamma_p$ and one can check that for $\gamma_p\sim 2\times 10^4$ the instability time-scale will no longer be small compared to the kinematic time-scale and the instability will be strongly suppressed. 

Fig. 2 shows the similar behaviour as in Fig. 1, but versus the relativistic factor of electrons. The set of parameters is the same as in Fig. 1, except $\gamma_p = 10^2$ and $\gamma_e = (2-2)\times 10^2$. Here as well, we see that for the mentioned physical parameters the instability is very efficient. Unlike the previous case, $\tau/P$ is a decreasing function of $\gamma_e$ and therefore, the higher its value, the more efficient the instability is.

In Fig. 3 we show $\tau/P$ versus the AGN luminosity. The parameters are the same as in Fig. 1 except $\gamma_e = 10^2$ and $L = (1-10)\times 10^{42}$ erg s$^{-1}$. On the plot we observe the similar situation, when the energy pumping process to Langmuir waves is very efficient and since the plot is a decreasing function of $L$ for more luminous AGNs the instability becomes stronger.

Fig. 4 shows the dependence of $\tau/P$ versus the black hole's mass. The parameters are the same as in the previous figure, except $L = 10^{42}$ erg s$^{-1}$ and $M_8 = 0.1-10$. It is evident that the instability time-scale is very small compared to the kinematic time-scale, therefore, the process is extremely efficient.

One important factor which can limit the process of amplification of Langmuir waves is Landau damping, characterised by the growth rate, $\Gamma_{_{LD}}\simeq n_{_{GJ}}\gamma_p^m\omega_p/(n_p\gamma_p)$ \citep{LD}. One can check that the LD increment is small compared to the wave excitation growth rate, implying that the electric fields' amplification will rapidly continue.

As a result, the electric fields' amplitudes will exponentially evolve in time
\begin{equation}
\label{E} 
E\propto e^{\Gamma t},
\end{equation}
which, for high values of $\Gamma$ might lead to pair creation. In particular, if the electric field reaches the value $1.4\times10^{14}$ statvolt cm$^{-1}$, efficient pair production will start via the Schwinger mechanism \citep{schwinger}. Therefore, a next step is to consider a similar problem which we have done for pulsars \citep{pair} and study pair cascading in AGN magnetospheres.

\section{Summary}

In the paper we considered the parametrical instability of energy pumping from rotation directly to the Langmuir waves by taking general-relativistic corrections to the GJ density.

To study this problem, we examined the set of linearised equations which fully describe the mentioned process: Euler equation, continuity equation and the Poisson equation and derived the dispersion relation of the Langmuir wave.

By solving the dispersion relation we obtained the instability growth rate which has been analysed versus several important physical parameters: electron's and proton's relativistic factors, AGN luminosity and the mass of the central black hole.

We have found that for a variety of cases the process of parametrically excited Langmuir waves is very strong,t leading to the exponential growth of the electrostatic field's amplitudes. 

It has been pointed out that such a behaviour might lead to pair cascading processes, which might be significant in AGN astrophysics.

\section*{Acknowledgments}
The work was supported by the EU fellowships for Georgian researchers.
Z.O. also would like to thank Torino Astrophysical Observatory and Universit\'a degli Studi di Torino for hospitality during working on this project.


\end{document}